\documentclass[12pt,preprint]{emulateapj}

\slugcomment{Received 2008 August 27; accepted by ApJL 2009 February 23}

\shorttitle{Accretion Variability}
\shortauthors{Nguyen et al.}

\begin{document}


\title{How Variable is Accretion in Young Stars?}


\author{Duy Cuong Nguyen\altaffilmark{1}, Alexander Scholz\altaffilmark{2}, Marten H. van Kerkwijk\altaffilmark{1}, Ray Jayawardhana\altaffilmark{1}, Alexis Brandeker\altaffilmark{3}}

\email{nguyen@astro.utoronto.ca}

\altaffiltext{1}{Department of Astronomy \& Astrophysics, University of Toronto, 50 St. George Street, Toronto, ON M5S 3H4, Canada; nguyen, mhvk, rayjay@astro.utoronto.ca}
\altaffiltext{2}{SUPA, School of Physics \& Astronomy, University of St. Andrews, North Haugh, St. Andrews, KY16 9SS, United Kingdom; as110@st-andrews.ac.uk}
\altaffiltext{3}{Department of Astronomy, Stockholm Observatory, SE-106 91 Stockholm, Sweden; alexis@astro.su.se}

\begin{abstract}
We analyze the variability in accretion-related emission lines for 40 Classical T~Tauri stars to probe the extent of accretion variations in young stellar objects. Our analysis is based on multi-epoch high-resolution spectra for young stars in Taurus-Auriga and Chamaeleon~I. For all stars, we obtain typically four spectra, covering timescales from hours to months. As proxies for the accretion rate, we use the H$\alpha$~10\%~width and the \ion{Ca}{2}-$\lambda$8662 line flux. We find that while the two quantities are correlated, their variability amplitude is not. Converted to accretion rates, the \ion{Ca}{2} fluxes indicate typical accretion rate changes of 0.35\,dex, with 32\% exceeding 0.5\,dex, while H$\alpha$~10\%~width suggests changes of 0.65\,dex, with 66\% exceeding 0.5\,dex. We conclude that \ion{Ca}{2} fluxes are a more robust quantitative indicator of accretion than H$\alpha$~10\%~width, and that intrinsic accretion rate changes typically do not exceed 0.5\,dex on timescales of days to months. The maximum extent of the variability is reached after a few days, suggesting that rotation is the dominant cause of variability. We see a decline of the inferred accretion rates towards later spectral types, reflecting the $\dot{M}$\ vs.\ $M$ relationship. There is a gap between accretors and non-accretors, pointing to a rapid shutdown of accretion. We conclude that the $\sim 2$ orders of magnitude scatter in the $\dot{M}$\ vs.\ $M$ relationship is dominated by object-to-object scatter instead of intrinsic source variability.
\end{abstract}


\section{Introduction}
\label{s1}

Young low-mass stars that are still accreting while contracting towards the main-sequence are identified observationally with T~Tauri stars \citep[e.g.][]{1989ARA&A..27..351B,1989A&ARv...1..291A}. One of their defining properties is variability \citep[e.g.][]{1891Obs....14...97K,1945ApJ...102..168J,1976ApJS...30..307R}, which can be traced to changes in continuum excess flux caused by varying accretion flow from a circumstellar disk \citep{1991ApJ...382..617H}; other possible sources of variability include cool spot rotation and extinction events. In addition to the continuum, this variability is seen in a number of emission lines, including the Hydrogen Balmer series, and has been attributed to rotation, variable accretion, and winds \citep[e.g.][]{1995AJ....109.2800J,2001ApJ...548..377B,2002ApJ...571..378A,2005A&A...440..595A}. Photometric monitoring campaigns for large numbers of objects have revealed that hot spots formed by gas accretion, co-rotating with the objects, are one of the most important sources of variability in young stars \citep[e.g.][]{1994AJ....108.1906H,1995A&A...299...89B,1996A&A...310..143F}. However, it remains unclear to what extent the photometric and spectroscopic variability seen in typical T~Tauri stars is directly related to changes in the accretion rate. As we argue below, this issue can be addressed by spectroscopic monitoring, but so far spectroscopic studies have been limited to small samples.

Inferences about accretion are usually made in the framework of the magnetospheric accretion scenario \citep[see the review by][]{2007prpl.conf..479B}. Recent observations have shown the presence of a correlation between inferred mass accretion rate and central object mass, extending over several orders of magnitude: $\dot{M} \propto M^{\alpha}$, with $\alpha \sim 2$ \citep[e.g.][]{2003ApJ...592..266M,2004A&A...424..603N,2005ApJ...626..498M}. Apart from this correlation, all $\dot{M}$ vs.\ $M$ plots feature a large scatter: at any given object mass, the accretion rates show a dispersion of about two orders of magnitude.

A number of ideas have been put forward to interpret these findings. The accretion rate vs. mass correlation has been attributed to a Bondi-Hoyle flow to the star-disk system \citep{2005ApJ...622L..61P}, to initial rotational velocities of collapsing cores \citep{Dullemond:2006p4130}, to a dispersion in disk parameters \citep{2006ApJ...639L..83A}, to a complex magnetic field geometry \citep{2006MNRAS.371..999G}, and to a declining disk ionization with stellar mass \citep{2003ApJ...592..266M,2005ApJ...626..498M}. Alternatively, it has been argued that the correlation is not physical, but rather reflects selection effects; therefore, the dominant feature is the scatter itself \citep{2006MNRAS.370L..10C}. All these scenarios predict specific values for $\alpha$ (0--2) and specific properties of the scatter.

A basic question about the $\dot{M}$\ vs.\ $M$ relation is whether scatter around it is due to object-to-object variations (i.e. the accretion rate for any individual object remains mostly constant) or due to a variable accretion rate in any given object \citep{2006ApJ...638.1056S}. At least for some objects, strong changes in accretion related lines have been reported, e.g.\ for the brown dwarf 2MASSW J1207334-393254 \citep{2005ApJ...629L..41S,2007ApJ...671..842S}, but it is not clear if such variations are common or not. Since most previous studies were based on single-epoch measurements of accretion rates, it was impossible thus far to address this issue.

Previous work has established that a number of optical emission lines originate in the accretion flow and are affected by the accretion rate. In particular, the H$\alpha$~10\%~width and the \ion{Ca}{2}-$\lambda$8662 line flux are found to correlate well with the accretion rate \citep{Muzerolle:1998p4143,2004A&A...424..603N,2005ApJ...626..498M}. These empirical indicators facilitate studies of accretion, and for the first time, allow us to investigate accretion in multi-epoch spectra for large samples of objects. Here, we use both diagnostics to probe the intrinsic accretion variability in young stars in Taurus-Auriga and Chamaeleon~I, to provide new observational limits for the aforementioned scenarios.

\section{Observations and Data Analysis}
\label{s2}

We obtained multi-epoch high-resolution optical spectra of $40$ members in the $\sim2$\,Myr old Chamaeleon~I and Taurus-Auriga star forming regions. The targets span the spectral types from F2 to M5 based on published classifications, and consist of accretors without suspected close companions selected from the list of \citet{Nguyen:2009}. The data were collected using the echelle spectrograph MIKE \citep{Bernstein:2003p626} on the Magellan Clay 6.5 meter telescope at the Las Campanas Observatory, Chile on 15 nights during four observing runs between 2006 February and 2006 December. Each target was observed typically at four epochs with baselines of hours, days, and months (only one target was observed just twice; 24, 12 and 3 targets were observed four, five and six times, respectively). For the data reduction, we used customized routines running in the ESO-MIDAS environment (described in detail in Brandeker et al., in preparation).

The H$\alpha$~10\%~widths were determined as follows. First, we estimated the continuum level by linearly interpolating flux values in the range of $500$ km~s$^{-1}$ to $1000$ km~s$^{-1}$ on either side of the line. Next, the maximum flux level of H$\alpha$ emission was measured with respect to this continuum level. Finally, the crossing points of the H$\alpha$ emission with the $10\%$ flux level were identified, and the width was measured. Sometimes absorption components in the H$\alpha$ emission line profile falls below the $10\%$ flux level. In such cases, we ensured that we measured the width consistently for all epochs, i.e. we measured widths to the edges of any blue- or red-shifted absorption features. We did not correct for underlying photospheric absorption in H$\alpha$.

We derived the \ion{Ca}{2}-$\lambda$8662 emission fluxes (${\mathcal F}_\textrm{\scriptsize{Ca}\,\tiny{I}}\!_\textrm{\tiny{I}}$) from the observed emission equivalent widths. To determine the widths, we integrated the emission above the continuum level. For emission profiles attenuated by a broad absorption feature, we used the median flux within 0.2\,\AA\ of the absorption minima as an approximate continuum level for integration, similar to what was done by \citet{Muzerolle:1998p4143}. To infer the emission fluxes from the equivalent widths, we must know the underlying photospheric continuum flux. We used the continuum flux predicted by the PHOENIX synthetic spectra for a specified $T_{\rm eff}$ and surface gravity. We inferred $T_{\rm eff}$ from our spectral types, and assumed a surface gravity of $\log g = 4.0$ (cgs units). In our estimate, we ignore veiling, which could lead to an underestimate of the line flux. Indeed, for five targets shared by \citet{2005ApJ...626..498M}, our results were lower by 0.05 to 0.41\,dex. We tried to measure veiling from our spectra, but we found the relatively poor S/N prevented us from reaching sufficient accuracy (S/N of our spectra was typically $\sim 20$ at H$\alpha$, whereas literature studies for veiling, e.g.\ \citet{1989ApJS...70..899H}, use spectra with S/N $\!\gtrsim\! 75$). However, for our purposes of studying variability, the bias due to veiling is not important.

\section{Accretion Indicators and Their Variability}
\label{s3}

\begin{figure}
\begin{center}
\includegraphics[width=8cm]{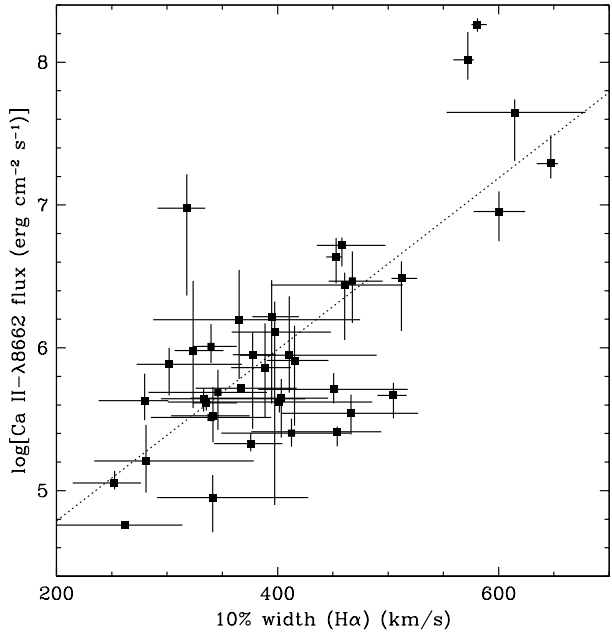}
\caption{\ion{Ca}{2}-$\lambda$8662 fluxes vs.\ H$\alpha$~10\%~widths in our sample: shown are the averages from our multi-epoch spectra with the error bars indicating the minimum-maximum range. From linear regression analysis, the probability that the data are not correlated is $<10^{-6}$; the best linear fit is overdrawn.
\label{f1}}
\end{center}
\end{figure}

Previous studies have shown that the H$\alpha$~10\%~width and the \ion{Ca}{2}-$\lambda$8662 flux are correlated with accretion rates determined using the traditional method based on optical veiling \citep{2004A&A...424..603N,2005ApJ...626..498M}, with \ion{Ca}{2} showing significantly less scatter around the fit \citep{2008ApJ...681..594H}. In Fig.~\ref{f1}, we compare these two accretion indicators for our sample, with the points set to the average values from the multi-epoch spectra, and the `error bars' indicating the range of values (the ranges are dominated by variability, not by measurement uncertainty). We find a clear linear correlation between the H$\alpha$~10\%~width and $\log$ flux in \ion{Ca}{2}-$\lambda$8662, with the deviations comparable to the scatter. This provides reassurance that both parameters are mainly determined by the same physical quantity, which we identify as the mass accretion rate based on published findings. For individual objects, the degree of correlation between the two accretion indicators varies, which may be explained by the varying extent to which accretion affects different emission lines. We do not see a significant correlation between the variability in the accretion indicators and stellar mass, Spitzer IR excess, or star forming region in our sample.

\begin{figure}
\begin{center}
\includegraphics[width=8cm]{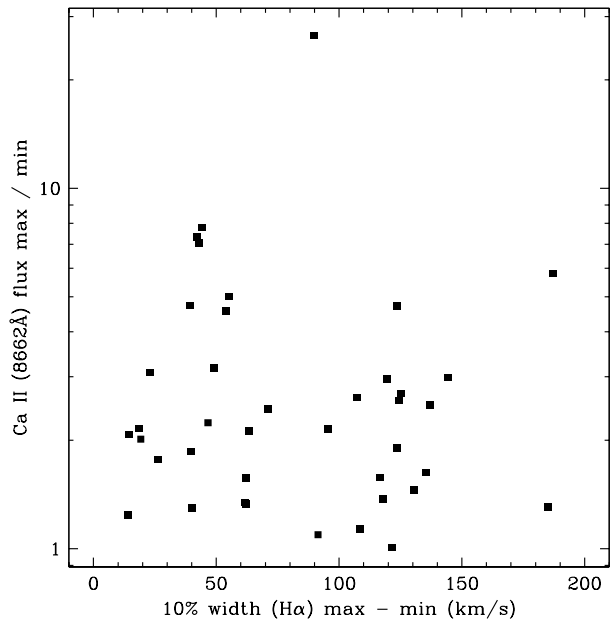}
\caption{Variability in spectroscopic diagnostics of accretion measured in our multi-epoch spectra for individual objects: $(\mathrm{max}/\mathrm{min})$ in \ion{Ca}{2}-$\lambda$8662 flux vs.\ $(\mathrm{max}-\mathrm{min})$ of the 10\%~width. Both quantities are supposed to be directly related to accretion rate changes. We do not find a significant correlation in this dataset (correlation coefficient $r^2 \sim 0.02$).
\label{f2}}
\end{center}
\end{figure}

In Fig.~\ref{f2}, we compare the variability in the two accretion indicators. Here we use $(\mathrm{max}-\mathrm{min})$ for the H$\alpha$~10\%~width and $(\mathrm{max}/\mathrm{min})$ for the \ion{Ca}{2}-$\lambda$8662 flux, because these quantities are directly proportional to accretion rate changes \citep{2004A&A...424..603N,2005ApJ...626..498M}. One sees that the data are not correlated (correlation coefficient $r^2 \sim 0.02$), indicating that at least one of these parameters does not reflect changes in the accretion rate.

Using the correlations $\log \dot{M} = 1.06 \log {\mathcal F}_\textrm{\scriptsize{Ca}\,\tiny{I}}\!_\textrm{\tiny{I}} - 15.40$ of \citet{2005ApJ...626..498M}, and $\log \dot{M} = -12.89 + 0.0097\,\textrm{H}\alpha\,10\%$ [km~s$^{-1}$] of \citet{2004A&A...424..603N}, we convert the variability seen in the spectroscopic indicators to accretion rate changes. Based on \ion{Ca}{2}-$\lambda$8662 fluxes, the median accretion rate variability in our spectra is 0.35\,dex, with 13/40 or 32\% exceeding 0.5\,dex and only 1/40 or 2.5\% exceeding one order of magnitude. These numbers should be treated as upper limits, as the uncertainties in measuring \ion{Ca}{2} fluxes likely contribute $\sim 0.1$\,dex to the scatter. In contrast, using the H$\alpha$~10\%~width gives a median variability of 0.65\,dex, with the majority exceeding 0.5\,dex (26/40 or 65\%), and still a substantial fraction exceeding one order of magnitude (16/40 or 40\%). 

This result indicates clearly that the \ion{Ca}{2}-$\lambda$8662 flux is a more robust quantitative diagnostic of accretion rate. Our empirical result is consistent with the models of Classical T~Tauri stars of \cite{Azevedo:2006p4136}, which find that \ion{Ca}{2} broad components are formed predominantly in the accretion flow, and track accretion rate, while the hydrogen lines are much more affected by stellar winds.

The main reason to use H$\alpha$ for accretion rate measurements is that it is much easier to measure, and is sensitive to lower levels of accretion. In comparison with line fluxes or equivalent widths, the 10\%~width also has the advantage of not being heavily affected by uncertainties in estimating the underlying continuum. This property is beneficial especially for faint objects: the typical variation in 10\%~width from continuum uncertainty is $\sim 15$ km~s$^{-1}$. Based on our findings, however, the 10\%~width should not be trusted for quantitative measurements. Nevertheless, it is still useful as a qualitative indicator of accretion.

A major problem in the 10\%~width measurement is its strong dependence on the line profile. The majority of the accreting stars in our sample show absorption components in their H$\alpha$ profile, caused by relatively cool gas, e.g. in stellar or disk winds, or parts of the accretion flow, seen in projection against the hot shock front. Since these absorption features are variable in intensity and in position relative to the emission, the maximum intensity and therefore the 10\% level can vary considerably from epoch to epoch. This effect can cause large changes in 10\%~width not necessarily associated with accretion rate changes.

\begin{figure}
\begin{center}
\includegraphics[width=8cm]{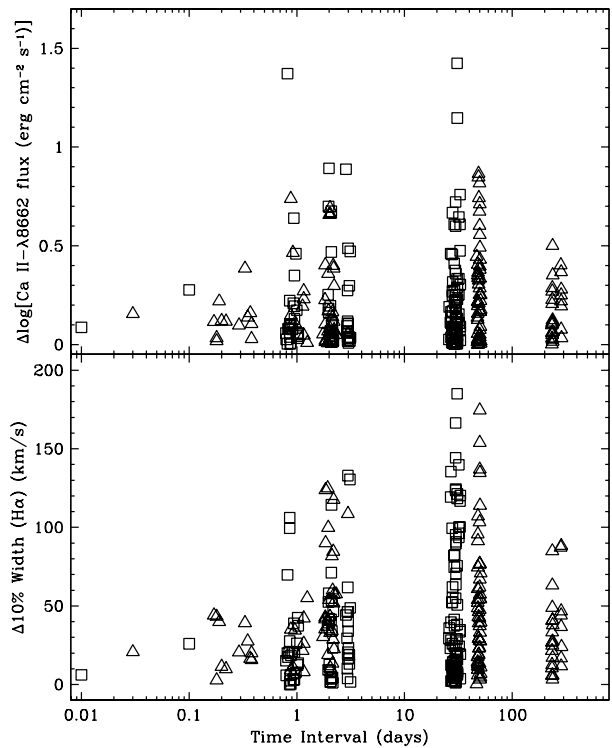}
\caption{Change in \ion{Ca}{2}-$\lambda$8662 flux (upper panel) and H$\alpha$~10\%~width (lower panel) vs. time. Data from Cha~I are represented by triangles, and those from Tau-Aur are drawn as squares. The maximum extent of the variability is reached after a few days and does not further increase towards longer timescales.
\label{f3}}
\end{center}
\end{figure}

To probe the timescales of variability in the two line indicators, we used our time series baselines of hours to months: we examined indicator changes between all combinations of epochs for each object, e.g. if an object was observed at epochs A, B \& C, then we reviewed changes between epochs A-B, A-C, and B-C. We show the changes graphically in Fig.~\ref{f3}. Both for the H$\alpha$~10\%~width and for the \ion{Ca}{2}-$\lambda$8662 flux, we obtain consistent results: the extent of the variability increases on timescales ranging from hours to several days. On longer timescales, the amount of variability saturates. Thus, the dominant timescale for accretion-related variability is several days. This idea suggests the variability is determined near the star, where the timescales are sufficiently short. Probably, one major factor is rotation, since typical rotation periods in young stars are in the range of 1--10\,d \citep{2007prpl.conf..297H}. In addition, there may also be a connection to the characteristic infall timescale, typically hours, or magnetic field reconnection events, perhaps several days. Note that we cannot probe long-term variations on timescales of years with our dataset. However, the photometric results of \citet{Grankin:2007p4162}, with observations secured over more than 20 years, find most Classical T~Tauri stars to be fairly stable on long timescales.

\section{The Origin of the Scatter in Accretion Rate\ vs.\ Mass}
\label{s4}

\begin{figure}
\begin{center}
\includegraphics[width=8cm]{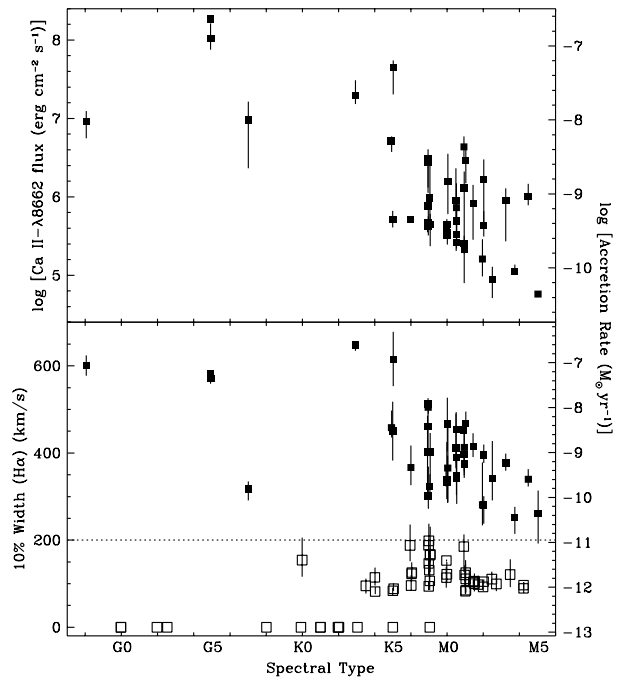}
\caption{\ion{Ca}{2}-$\lambda$8662 flux (upper panel) and H$\alpha$~10\%~width (lower panel) vs.\ spectral types. Error bars indicate the full range of the variations. In the lower panel, we also show the non-accretors (hollow symbols) with H$\alpha$~10\%~widths $<200$\,km~s$^{-1}$ (after subtracting rotational broadening). Objects without H$\alpha$ emission are shown with a 10\%~width of $0$\,km~s$^{-1}$. The right axis shows the accretion rates inferred from the relations from \citet{2005ApJ...626..498M} and \citet{2004A&A...424..603N}. For the M and K type stars, one sees a clear rise in H$\alpha$~10\%~width towards earlier spectral type, but at K0 and earlier, it appears to plateau, with no objects showing widths larger than $\sim 600$\,km~s$^{-1}$.
\label{f4}}
\end{center}
\end{figure}

Our new constraints on accretion variability in a large sample allows us to constrain the origin of the distribution of data points in the accretion rate vs.\ mass diagram. In Fig.~\ref{f4}, we show the two accretion indicators used in this study as a function of spectral type. For the H$\alpha$~10\%~width, we also include the non-accreting objects. In both cases, a clear trend is seen: the spectroscopic indicator drops towards later spectral types, which reflects the $\dot{M}$ vs.\ $M$ correlation reported in previous studies (see \S\ref{s1}). As can be seen in the H$\alpha$~10\%~widths, the opposite trend is seen for the non-accretors, which is likely due to increasing levels of magnetic activity towards later spectral types. This leads to a `U-shaped' distribution of data points: accretors and non-accretors are separated by a large gap at F--K spectral types, but are harder to distinguish in M-type objects.

As can be seen in both panels, the total amount of scatter in the accretion indicator cannot be explained by intrinsic accretion rate variability, which is shown as error bars. This is particularly obvious from the \ion{Ca}{2}-$\lambda$8662 fluxes, which we identified as the more robust accretion indicator in \S\ref{s3}. Converted to accretion rates, the objects cover 1--2 orders of magnitude at any given spectral type, while the variability accounts only for a small fraction of this scatter ($\sim 0.35$\,dex, see \S\ref{s3}). In other words, on timescales of days to months the objects do not move significantly in the diagram.

Although we cannot exclude the presence of significant variability on longer timescales, the scatter in the $\dot{M}$ vs.\ $M$ plot is most likely dominated by object-to-object variations. This confirms previous claims by \citet{2004A&A...424..603N} and is contrary to suggestions by \citet{2006ApJ...638.1056S} based on a much smaller sample. Thus, stellar mass or parameters that strictly scale with stellar mass are clearly not the only factors that affect the accretion rate. 

One possible explanation for the spread of accretion rates is the different evolutionary stages of the objects considered here. Accretion rates are expected to drop with age, either following a power law, if the timescale is determined by the slow viscous evolution of the disk \citep{2006MNRAS.370L..10C}, or in a more rapid process, consistent with the rapid inner disk clearing inferred from the scarcity of `transition' objects with optically thin inner disks. The presence of the clear gap between the accreting and non-accreting populations in Fig.~\ref{f4}, particularly well-defined for the higher mass objects is thus highly interesting. The scarcity of objects in the transition from accretors to non-accretors argues for a rapid evolution between states, as suggested in disk evolution models including mass loss from the disk \citep[e.g.][]{2001MNRAS.328..485C,2003MNRAS.342.1139A}. This may be related to the similarly rapid timescale of transition between objects with IR excess and those without.
 
Apart from the evolutionary stage, a number of other factors have been suggested to influence the accretion rates. If disk ionization is a major factor, our findings imply that the main source of ionization is probably external (e.g. cosmic rays), as the stellar ionising radiation is expected to scale with mass and thus would not allow for a wide range of accretion rates at constant mass. In general, our results favour scenarios where initial or environmental conditions are more important than stellar parameters. For instance, a model where the dispersion in initial disk parameters, e.g.\ mass and radius, determines the distribution of data points in the $\dot{M}$ vs.\ $M$ plot, as recently described by \citet{2006ApJ...639L..83A}, would be consistent with our results.

To constrain the influence of stellar and environmental parameters on the appearance of the $\dot{M}$ vs.\ $M$ plot, it would be necessary to isolate them from evolutionary effects. It would be of particular interest to investigate accretion rates versus other parameters along lines of constant mass in the $\dot{M}$ vs.\ $M$ diagram. Unfortunately, for these tasks the currently available samples may not be sufficiently large. In principle, however, such analyses have the potential to provide important information on the nature of the accretion process in young stars. Finally, the causal relationships between our accretion indicators and accretion rates are not fully understood, and we use them merely as empirical tools, based on published correlations. Nevertheless, the fact that indicators such as \ion{Ca}{2}-$\lambda$8662 line flux correlate remarkably well with the current empirical framework of measuring accretion rates gives some confidence that these indicators relate to physical reality.

\acknowledgments

We thank the anonymous referee for a very helpful review and constructive critical comments that greatly improved the clarity of the letter. DCN acknowledges gratefully the hospitality of the astronomy group in St. Andrews during his visit. This work was supported in part by NSERC grants to RJ and MHvK and an Early Researcher Award from Ontario to RJ.

\end{document}